\documentclass[preprint2]{aastex62}
\usepackage{graphicx}
\usepackage{mathtools}
\usepackage{xcolor}
\usepackage{booktabs}
\usepackage{color,soul}
\usepackage{amsmath}
\usepackage{hyperref}
\usepackage{times}

\begin{document}

\title{The Impacts of Modeling Choices on the Inference of Circumgalactic Medium Properties from Sunyaev-Zeldovich Observations}

\author{Emily Moser}
\affiliation{Department of Astronomy, Cornell University, Ithaca, NY 14853, USA}

\author{Stefania Amodeo}
\affiliation{Department of Astronomy, Cornell University, Ithaca, NY 14853, USA}
\affiliation{Universit\'{e} de Strasbourg, CNRS, Observatoire astronomique de Strasbourg, UMR 7550, F-67000 Strasbourg, France}

\author{Nicholas Battaglia}
\affiliation{Department of Astronomy, Cornell University, Ithaca, NY 14853, USA}

\author{Marcelo A. Alvarez}
\affiliation{Lawrence Berkeley National Laboratory, One Cyclotron Road, Berkeley, CA 94720, USA}
\affiliation{Berkeley Center for Cosmological Physics, UC Berkeley, CA 94720, USA}

\author{Simone Ferraro}
\affiliation{Lawrence Berkeley National Laboratory, One Cyclotron Road, Berkeley, CA 94720, USA}
\affiliation{Berkeley Center for Cosmological Physics, UC Berkeley, CA 94720, USA}

\author{Emmanuel Schaan}
\affiliation{Lawrence Berkeley National Laboratory, One Cyclotron Road, Berkeley, CA 94720, USA}
\affiliation{Berkeley Center for Cosmological Physics, UC Berkeley, CA 94720, USA}

\begin{abstract}
As the signal-to-noise of Sunyaev-Zeldovich (SZ) cross-correlation measurements of galaxies improves our ability to infer properties about the circumgalactic medium (CGM), we will transition from being limited by statistical uncertainties to systematic uncertainties. Using thermodynamic profiles of the CGM created from the IllustrisTNG (The Next Generation) simulations we investigate the importance of specific choices in modeling the galaxy sample. These choices include different sample selections in the simulation (stellar versus halo mass, color selections) and different fitting models (matching by the shape of the mass distribution, inclusion of a two-halo term). We forward model a mock galaxy sample into projected SZ observable profiles and fit these profiles to a generalized Navarro-Frenk-White profile using forecasted errors of the upcoming Simons Observatory experiment. We test the number of free parameters in the fits and show that this is another modeling choice that yields different results. Finally, we show how different fitting models can reproduce parameters of a fiducial profile, and show that the addition of a two-halo term and matching by the mass distribution of the sample are extremely important modeling choices to consider.
\end{abstract}

\section{Introduction}

The circumgalactic medium (CGM) consists of a large reservoir of gas surrounding the outermost regions of galaxies. It is an important component of the galactic structure, and is believed to govern the cyclical flow of material in and out of galaxies. This cycle includes material falling onto the disks of galaxies from the CGM and intergalactic medium (IGM) and being recycled back out through various feedback mechanisms. 

The CGM is known to be complex and multiphase in temperature, ionization state, and kinematics \citep{Tumlinson2011,Werk2013,Nielsen2015,Tumlinson2017,Oppenheimer2018} and has been studied through observations (e.g. \citet{Lanzetta1995,Chen1998,Steidel2010,Tumlinson2011,Rudie2012,Tumlinson2013,Werk2014})
and hydrodynamic simulations (e.g. \citet{Oppenheimer2008,Ford2013,Hummels2017,Suresh2017, Oppenheimer2018,Hummels2019,Peeples2019}). Several questions regarding the physical processes that govern the different states of the CGM remain unanswered, commonly including 
what are the properties of quenched galaxies and how do the galaxies become quenched \citep{Baldry2004,Noeske2007,Peng2010,Geha2012}, and the \textit{missing baryon problem} in which baryons predicted by the current cosmological model are not observed \citep{Persic1992,Fukugita1998,Peebles2004,Cen2006,Bregman2007,Dave2009}. Star formation in galaxies is known to be inefficient \citep{Peebles2004,Federrath}, which implies that there is some process occurring within the CGM and its interaction with the host galaxy that is preventing the gas from cooling and forming stars efficiently. This process could be related to the various feedback mechanisms injecting energy into the system such as supernovae and active galactic nuclei (AGN; \citet{DiMatteo2005,Springel2005,Scannapieco2008,Somerville2008,Marasco2015,Somerville2015,Agertz2016} that push central material to the far parts of the galaxy while also affecting its thermodynamic properties. These problems and many more discussed in the literature are important to understand for galaxy evolution theory, and can be studied theoretically through cosmological simulations and observationally through galaxy surveys at multiple redshifts and wavelengths.

An emergent method to observe the CGM is observing the cosmic microwave background (CMB). While initially measured to be homogeneous and isotropic, fluctuations (called anisotropies) in temperature have been detected, and these anisotropies can give insight on the seeds of structure formation and growth. Secondary anisotropies due to the Sunyaev-Zeldovich (SZ) effect \citep{SZ1970} have also recently been measured, which can be further broken down into thermal (tSZ) and kinetic (kSZ) effects \citep{SZ1972, SZ1980}.

The tSZ effect describes the increase in energy of CMB photons due to scattering off ionized electrons in galaxies and galaxy clusters, and produces distortions in the nearly perfect blackbody spectrum of the CMB. It is a function of frequency, $\nu$, and the Compton-y parameter, shown in Equation~\ref{eq:tSZ}. The amplitude of this distortion is proportional to the line-of-sight (LOS) integral of the electron pressure, so it can be used as a probe of the gas pressure within galactic halos. 
\begin{equation}
    \begin{gathered}
    \frac{\Delta T(\nu)}{T_{\text{CMB}}} = f(\nu)y(\theta) \, , \\ 
    y(\theta)= \frac{\sigma_T}{m_{e}c^2}\int_{\text{LOS}}P_{e}(\sqrt{l^2+d_A^2(z){|\theta|}^2})dl \label{eq:tSZ}
    \end{gathered}
\end{equation}
where $\Delta T(\nu)$ is the shift in temperature measured as the tSZ signal, $T_{CMB}$ is the temperature of the CMB, spectral function $f(\nu) = x\text{coth}(x/2)-4$, with $x=\frac{h\nu}{k_{B}T_{\text{CMB}}}$,  h is the Planck constant, $k_B$ is the Boltzmann constant, $y(\theta)$ is the Compton-y parameter measured within $\theta$, $\sigma_T$ is the Thomson scattering cross section, $m_e$ is the electron mass, c is the speed of light, $d_A(z)$ is the angular diameter distance at redshift $z$, and $P_e$ is the electron pressure.

The kSZ effect is the Doppler shift of CMB photons scattering off free electrons in galaxies and clusters with peculiar velocities, causing Doppler shifts in the CMB temperature that are directly related to the peculiar momentum. These shifts in temperature are proportional to the LOS integral of peculiar velocity multiplied by electron number density; therefore, we can use the kSZ to measure the density of the CGM in extragalactic halos \citep{Battaglia2017}.
\begin{equation}
    \begin{gathered}
    \frac{\Delta T}{T_{\text{CMB}}} = \frac{\sigma_T}{c}\int_{\text{LOS}}e^{-\tau(\theta)}n_{e}v_{p}dl \, , \\ \tau(\theta)=\sigma_T \int_{\text{LOS}}n_e(\sqrt{l^2+d_A^2(z){|\theta|}^2})dl \label{eq:kSZ}
    \end{gathered}
\end{equation}
where $n_e$ is the electron number density, $v_p$ is the peculiar velocity, and $\tau(\theta)$ is the optical depth.

Combining the tSZ and kSZ effects provides complete thermodynamic information of the CGM \citep{Battaglia2017} and can provide constraints on the physical processes that govern star formation and feedback, and thus galaxy evolution. The combination of SZ measurements is a relatively new method to study the CGM. While the tSZ effect has been observable in galaxies and galaxy clusters for several years, detections of the kSZ effect have only recently been possible through cross-correlation of CMB observations with galaxy catalogs (e.g. first detected in \citet{Hand2012}). Among the kSZ estimators that exist in the literature, a recent powerful method uses a velocity-weighted stack, in which the signal measured for each halo on the CMB map is weighted by an estimate of its LOS velocity, reconstructed from the density field \citep[e.g.][]{Planck_kSZ2016,Schaan2016, Schaan2021}. Using this method, recent results from the Atacama Cosmology Telescope (ACT) DR5 and Planck \citep{Amodeo2021,Schaan2021} have achieved high signal-to-noise measurements of the kSZ and tSZ effects, and in turn, of the electron density, temperature, and pressure distribution around the CMASS (constant stellar mass) galaxies from the Baryon Oscillation Spectroscopic Survey \citep[BOSS;][]{Ahn2014},  constraining the effects of feedback and finding tensions with cosmological simulations.
The signal-to-noise ratio ($S/N$) in measurements like those of \citet{Schaan2021} and \citet{Amodeo2021} will increase rapidly, as forecasted in \citet{Battaglia2017}, and will make the tSZ and kSZ effects even more powerful probes of astrophysical processes related to galaxy formation and evolution \citep{SO2019,CMB-S4-SRD,BH2019}. With such a high $S/N$, correctly modeling details of the sample becomes very important, and this is the crucial question that this study addresses.

Since it is not possible to directly observe galaxy evolution over cosmological timescales, simulations offer the only method of tracking and characterizing the physical properties and locations of baryons in the galaxies (including the CGM and even farther into the IGM) in controlled environments at multiple redshifts, which can act as predictors for how we believe real galaxies behave. In this study we have developed methods using cosmological simulations to study the thermodynamic properties of galaxy halos, thus constraining the baryonic processes that are responsible for making star formation inefficient and affecting the galaxy's evolutionary track.  

In the current era of high signal-to-noise SZ observations, modeling the halo samples becomes a major and important uncertainty to understand. Not taking certain properties of the sample into account in the model could bias the way in which the observations are interpreted. 

In this study we use simulated halo samples from the IllustrisTNG (The Next Generation) simulation \citep{Marinacci2018,Naiman2018,Pillepich2018,TNG,Nelson2018,Nelson2019} to test various components of modeling an example of an observed halo population. We model a subsample of the CMASS sample from the BOSS survey, Data Release 10 \citep{Ahn2014}, of which SZ profile measurements have already been made \citep{Amodeo2021,Schaan2021}. This subsample is chosen to contain galaxies in the region covered by ACT, that are mainly central galaxies of group-sized halos, selected with a halo mass lower than $10^{14} M_\odot$. The sample is spectroscopic, with a range of redshifts $0.4 < z < 0.7$ and median redshift $z=0.55$. This is just one example of a possible observed halo sample, but the results are expected to be applicable to modeling other samples as well. 

In this paper we summarize the simulations used in Section~\ref{sec:sims}. We discuss how we extract the simulated halo information and construct thermodynamic profiles in Section~\ref{sec:illstack}, the modeling uncertainties explored in Section~\ref{sec:modeling}, and the fitting procedure in Section~\ref{sec:fitting}. We discuss our methods of modeling the CMASS sample in Section~\ref{sec:modeling_2d}, and projecting the simulated three-dimensional profiles into a two-dimensional, observing space in Section~\ref{sec:projections}. Finally, we discuss the results of fitting in 3D in Section~\ref{sec:results_3d} and 2D in Section~\ref{sec:results_2d}, and how the modeling uncertainties affect the shape of the profiles and fitting parameters in each.

\section{Theoretical Profiles}\label{sec:3d}
We begin by describing our methods for producing and analyzing simulated three-dimensional profiles, which will also be used for computing and analyzing the two-dimensional profiles discussed in Section~\ref{sec:methods_2d}.

\subsection{Simulations}\label{sec:sims}
We have developed methods using the Illustris and IllustrisTNG simulations to study halo properties. Illustris \citep{Illustris} is the original large-scale hydrodynamical simulation of galaxy formation, and provides the foundation on which its successor TNG \citep{TNG} was built. Illustris includes models for gas cooling, stellar evolution, various forms of feedback (e.g. from supernovae and AGN), and many other physical processes to produce galaxy and cluster samples that matched observed trends and scaling relations well \citep{Illustris}. However, Illustris did struggle to reproduce other trends, such as low gas mass fractions in halos, discussed in detail in \citet{Genel2014,Illustris,Nelson2015}. Thus, the TNG simulations were created to address these discrepancies, along with including numerical advancements to the overall simulation code. TNG includes the same basic physical processes as the Illustris simulations, and additionally includes a new implementation of growth and feedback from supermassive black holes (SMBHs), galactic winds, and magnetic fields based on ideal magnetohydrodynamics \citep{Weinberger2017,Marinacci2018,Pillepich2018}.\footnote{We note that TNG uses the Planck cosmology from \citet{PlanckCollab2016}, while Illustris adopted the Wilkinson Microwave Anisotropy Probe cosmology from \citet{Hinshaw2013}.}

Both Illustris and TNG simulations have varying box sizes and resolutions. For Illustris, the smallest volume and lowest resolution run is Illustris-3, which has a box size of $106.5^3 {{\rm Mpc}}^3$ and tracks $2\times455^3$ gas and dark matter particles. For TNG the smallest volume and lowest resolution run currently available is TNG100-3, with a box size of $110.7^3$ ${{\rm Mpc}}^3$ and $2\times455^3$ tracked particles. The TNG simulation with the next lowest level of resolution for the same box size is TNG100-2, and has $2\times910^3$ resolution elements. 
In the following section, we compare the thermodynamic profiles of halos generated from these sets of simulations and we focus on TNG100-3 for the remainder of the paper

\subsection{Generating 3D Profiles}\label{sec:illstack}
We have identified simulated halos with certain properties of interest and created thermodynamic radial profiles using the publicly available stacking code repository \texttt{Illstack}.\footnote{\url{https://github.com/marcelo-alvarez/illstack}} An already existing public TNG repository \texttt{illustris-python}\footnote{\url{https://github.com/illustristng/illustris_python}} allows for the extraction of particle information from each snapshot, and offers access to the group catalogs of the simulations which include positions, masses, radii, and other quantities of the halos defined by the ``friends-of-friends" algorithms. In theory, one could study the simulated halo properties only using the information provided in the group catalog, however the study would be limited by the predetermined cuts and boundaries defined by the TNG halo finding algorithms in post-processing and would lack particles beyond a certain radius.  \texttt{Illstack} expands on the framework of \texttt{illustris-python} by storing the quantities of interest of every particle in the entire snapshot along with the halo positions of the group catalog, then linking the particles to the halos out to arbitrarily high radii. We do not perform cuts in the phase space of the simulated gas particles; rather, we select the halos for which we compute profiles by their group properties (mass and/or color) discussed in Section~\ref{sec:modeling}. The halos are then split into several radial bins, and the quantities are stacked in each of these bins to create 3D radial profiles. We define the radial bins to extend from $1\rm\times 10^{-4} - 1\rm\times 10^2$ Mpc split into 25 bins in logspace, resulting in radial bin size $\sim 0.55$ Mpc. \texttt{Illstack} also accounts for periodic boundary conditions to ensure that all halo and particle information is utilized even if a halo is located near the edge of the simulated box. 

\begin{figure*}
	\centering
	\includegraphics[scale=0.55]{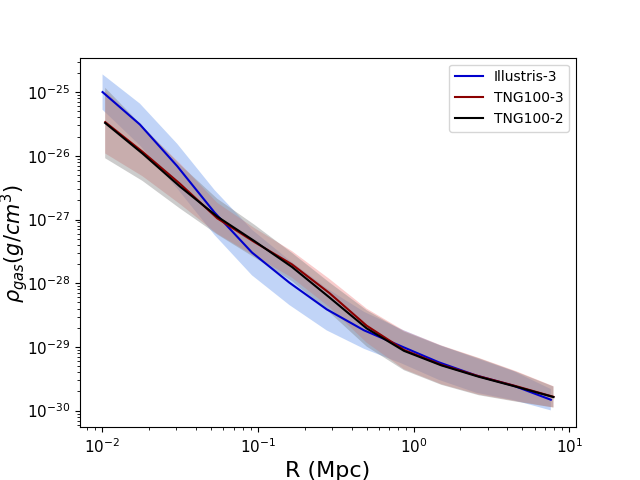}
	\includegraphics[scale=0.55]{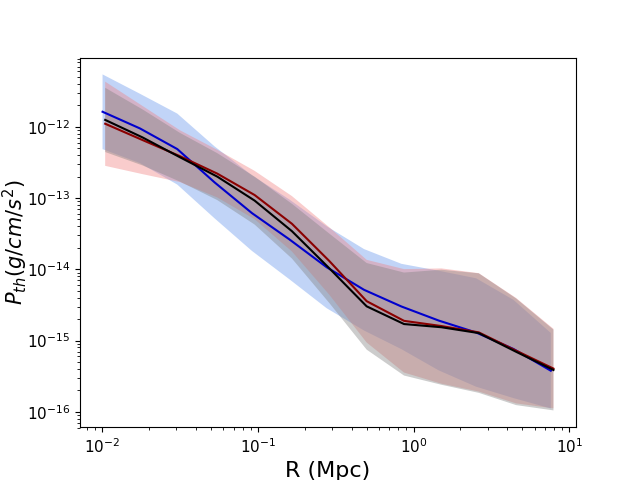}
	\vspace{-0.3cm}
	\caption{\label{fig:prof} Mean profiles of 3D gas density (left) and thermal pressure (right) of Illustris (blue), TNG100-3 (red), and TNG100-2 (black) as functions of radius at redshift $z = 0.55$ for halo masses $12 \leq \log_{10}(M_h/M_{\odot}) \leq 13$. The bands show the $\pm 1\sigma$ distribution of the values in each radial bin.}
\end{figure*}

Examples of the output from \texttt{Illstack} are shown in Figure~\ref{fig:prof}. In particular, the profiles display the properties of pressure and gas density to be directly related to the observable tSZ and kSZ effects, respectively. This figure shows the density and pressure profiles derived from halos of the same mass range ($12 \leq \log_{10}(\frac{M}{M_{\odot}}) \leq 13$, with $M$ as the halo mass of the group catalog) and same redshift ($z=0.55$) of the three different simulations described previously in Section~\ref{sec:sims}.
The solid lines show the median of each radial bin, and the bands show the $\pm 1\sigma$ distribution of the values within each radial bin. Differences among all of these simulations can be seen in the profiles, such as higher density values at inner regions and lower density values from middle regions from Illustris compared to TNG, and similar differences in the pressure profiles, which could be a result of the different prescription of AGN feedback among the simulations. Since the shortcomings of Illustris are well known and documented, we focus our analyses on TNG for the remainder of the paper.

Furthermore, given the mass ranges in which we are interested we do not see significant differences between the resolution levels of the TNG simulations, and thus do not require the higher resolution. Therefore we can be efficient with our computing resources and use the smallest box currently available with the lowest resolution, TNG100-3. 

\subsection{Modeling Uncertainties}\label{sec:modeling}
Once the simulated profiles are computed, we can further divide them to study various halo sample modeling uncertainties using \texttt{Illstack}. It has the ability to select and make profiles for halos of different types (e.g. within a desired mass range, different redshifts, and colors). We test different options for modeling halo populations including splitting by mass and color, adding a two-halo term to the fitting model, and matching the shape of an observed mass distribution. 

\subsubsection{Mass-splitting Definitions}\label{sec:mass_splits}
Relating the stellar contents of a galaxy to the dark matter halo in which it resides has been the subject of many studies in the past decade (e.g., \citet{Leauthaud2011,Leauthaud2012,Behroozi2010,Behroozi2013}), and an understanding of the topic is essential for understanding how dark matter plays a role in galaxy formation and evolution. This widely studied relation is called the stellar-halo mass relation (SHMR), and it attempts to map the stellar mass of a galaxy, usually estimated through photometric or spectroscopic surveys, to the underlying dark matter halo mass, usually calculated through lensing, kinematics, or abundance matching.
The functional form of the relation is debated, as well as its redshift dependence \citep{Behroozi2013}. 

As described in \citet{Behroozi2010}, measurements of the stellar mass of galaxies introduce significantly more systematic uncertainties than measurements of the halo mass, which could be a reason the computed relations vary among studies using different surveys and methods \citep{Shankar2014}. While the uncertainty in stellar mass function is more troublesome, there are still systematics in the derivation of halo masses, described in more detail by \citet{Behroozi2010}. With all of the uncertainties in both sides of the relation, the resulting relation is therefore quite uncertain. 

We use the TNG simulations in this study so that we have values for both stellar and halo masses. \texttt{Illstack} has the ability to select for halos by either mass option, so we derive stacked profiles for two separate populations based on mass type. This allows us to explore the differences in profiles of these samples, along with effects of stacking on samples with different mass distributions discussed further in Section~\ref{sec:3d_weighting}, we show how important it is to correctly model the sample, including the SHMR. Our exploration is specific to TNG, but it does provide a range of uncertainty that could arise from the SHMR.

Aside from looking at differences in the samples based on the mass type selection, we also perform a study on how the profiles change as a function of mass (both stellar and halo). We derive the TNG samples to match the limits of the CMASS sample analyzed in \citet{Amodeo2021} and discussed further in Section~\ref{sec:modeling_2d}). We further explore any trends in mass by splitting this sample into four smaller mass bins, discussed further in Section~\ref{sec:results_mass_dependence}.

\subsubsection{Color-splitting Definitions}\label{sec:color_splits}
One of the previously mentioned enigmas of CGM theory is the cause of galaxy quenching. The theoretical picture for quenched galaxies is that they are mostly red in color, as a result of little or no star formation. With \texttt{Illstack} we can separate the simulated halos into a red population and study the differences in the equivalent profiles of the entire sample that also includes active, bluer halos. Since the group catalogs of the Illustris simulations do not provide color information for the main halo, we use the colors from the most massive subhalo within the host halo of the TNG group catalog as representative of the whole halo. We derive subpopulations within each of the mass bins based on color using the TNG color cut in \citet{Nelson2018}, which is defined as the difference in magnitudes $g-r > 0.6$.  Using this color cut, we split the populations into one without a color cut (`tot' for total), and one with a red color cut (`red').

\subsubsection{Two-halo Term}\label{sec:two_halo}
Since \texttt{Illstack} can make profiles out to arbitrarily high radii, it can be used to study other theoretical aspects of the interaction of the CGM and IGM, i.e. the two-halo term. As can be seen in Figure~\ref{fig:prof} the profiles are generally decreasing functions of radius but there is a point of inflection near the outer part of the profile. This part of the profile has been called the \textit{two-halo term} and is due to contributions from neighboring halos and the overall enhancement of density and pressure from gravitational clustering of gas that has not yet virialized \citep{Cooray2002,Vikram2017,Hill2018}.

\begin{figure}
    \centering
    \includegraphics[scale=0.55]{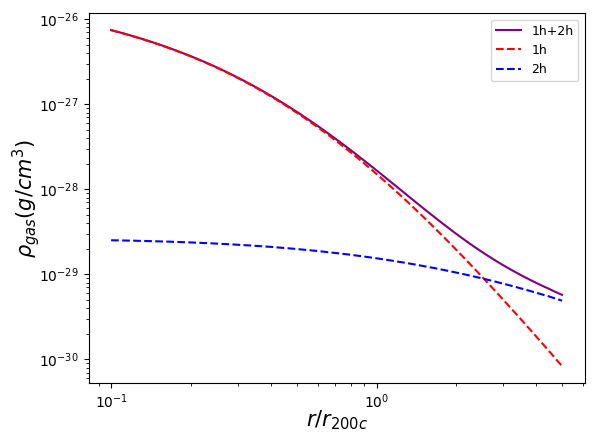}
    
    \includegraphics[scale=0.55]{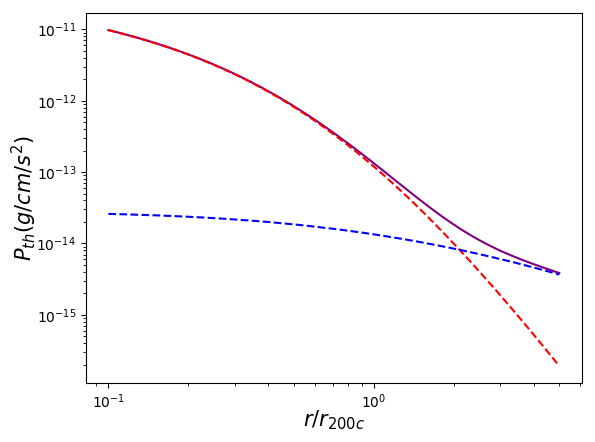}
    \caption{Density profile (top) and pressure profile (bottom) of a halo with mass $2\times{10}^{13}M_{\odot}$ and redshift $z=0.57$ as functions of scaled radius. The profiles are split into the components of the profiles as a function of scaled radius for the one-halo (red, dashed) and two-halo (blue, dashed) terms, using the two-halo term model of \citet{Vikram2017}. The profile combining the terms is shown in purple.}
    \label{fig:twohalo}
\end{figure}

The part of the profile associated with virialized gas in the main halo is called the ``one-halo term" and represents the signal only from the main halo. A model for the two-halo term was first described by \citet{Vikram2017}, but it is largely neglected in SZ measurement literature. It has been shown that the two-halo term contribution to measured halo signals can be significant, especially around lower mass halos \citep{Vikram2017,Hill2018}, and is therefore important to include when modeling observed signals. We show the contributions of the different terms as a function of radius in Figure~\ref{fig:twohalo}. The figure shows a density profile (top panel) and pressure profile (bottom panel) for a halo of mass $2\times{10}^{13}M_{\odot}$ at redshift $z=0.57$ as functions of radius scaled by the virial radius, $r_{200c}$, defined as the radius of the sphere whose mean density is 200 times the critical density of the universe at redshift $z$. The red and blue dashed lines correspond to the one-halo and two-halo terms, respectively, and the solid purple line shows the total profile combining the terms. It can be seen that for this halo, the two-halo term begins to significantly contribute to the profile around 1 virial radius. The amplitude and radius at which the two-halo term begins to contribute varies according to the mass and redshift of a halo, discussed further in \citet{Vikram2017}.

Here we include the model for the two-halo term for both density and pressure profiles derived in \citet{Vikram2017}, multiplied by an amplitude $A_{2h}$ that we allow as a free parameter when fitting the profiles as in \citet{Amodeo2021}, discussed further in Section~\ref{sec:fitting}. 

\subsubsection{Matching the Shape of the Mass Distribution}\label{sec:3d_weighting}
\begin{figure*}
	\centering
	\includegraphics[scale=0.55]{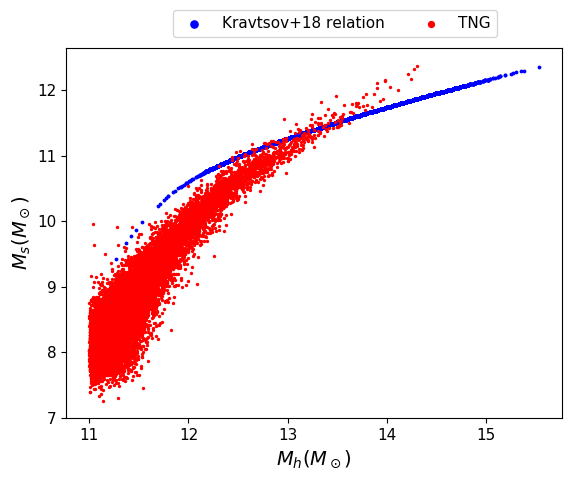}
	\includegraphics[scale=0.55]{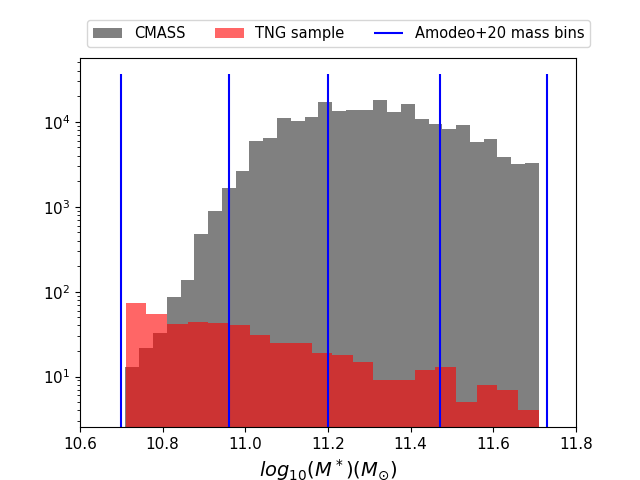}
	\vspace{-0.3cm}
	\caption{\label{fig:msmh_relation_and_cmass} Left panel: Comparison of the SHMR used for calculating corresponding halo masses for CMASS stellar mass sample \citep{Kravtsov2018} and SHMR of TNG. The SHMR relation compared to the simulations sample is clearly different, showing the importance of this modeling uncertainty, which we explore through modeling the CMASS sample two ways, via halo masses and stellar masses. Right panel: Histograms of the CMASS sample analyzed in \citet{Amodeo2021} shown in gray and the TNG mh sample in red. The blue vertical lines show the weights derived by \citet{Amodeo2021}. This comparison shows that while the limits of the TNG sample are chosen to match those of the CMASS sample, the distributions are not the same shapes, resulting in the importance of considering distribution matching.}
\end{figure*}

The distribution of the halos in the simulation does not match the distribution of the data as seen in the right panel of Figure~\ref{fig:msmh_relation_and_cmass}, therefore a straight average of the halos in the mass range could be biased. We test the importance of matching the shape of the mass distribution by weighting each halo appropriately to ensure that the contribution of each mass bin of the simulated sample matches the contribution of the corresponding mass bin of the observed sample.

We use the same weights as those used in \citet{Amodeo2021}. In more detail, the weights are computed so the area of the histogram of the CMASS sample (shown in the right panel of Figure~\ref{fig:msmh_relation_and_cmass}) is equal to 1 (i.e. $\sum_i w_i\Delta m = 1$, where $w_i$ are the weights of each mass bin and $\Delta m$ is the width of the bin). The bin edges are shown by the blue vertical lines in the right panel of Figure~\ref{fig:msmh_relation_and_cmass}, so a different weight is applied to the halos falling in each of these bins, according to their mass. This is especially important for the TNG halos that fall into the first two mass bins, since the shape of the distribution within these bins significantly differs from the observed CMASS distribution. We show this can cause considerable bias in the inferences of the profiles' parameters.

The specific method for the weighting is we assign each halo a weight according to its mass, then do a weighted average for all halos within each radial bin to get the \textit{matched} stack. The equation for the matched profile value in each radial bin would be ${val_{m}} = \frac{\sum val*w}{\sum w}$. In contrast, the unmatched (unweighted) average would be ${val}_{um} = \frac{vals}{N}$, ($N$ is number of halos in the bin), with all halos contributing an equal amount. This process is only relevant for the calculation of the raw 3D radial profiles; we are not weighting the observed 2D profiles, rather we do the weighing and stacking of the 3D profiles then project the average (weighted and unweighted) into the 2D SZ observables.

As the tSZ signal is proportional to $M^{5/3}$ and the kSZ signal is linearly dependent on mass (see Equations~\ref{eq:tSZ} and~\ref{eq:kSZ}), we expect the distribution matching to have more of an effect on the pressure profiles than density. However, we provide both matched and unmatched results for both quantities. This kind of weighting could be done for the redshift distribution as well, but as described in \citet{Amodeo2021}, the CMASS redshift distribution is peaked around the median and the resulting profiles with redshift weighting do not significantly differ from profiles without redshift weighting. Therefore, we simply use the median redshift of the CMASS distribution.

We also explore the effects of the different mass type-selected samples to further study the importance of correctly modeling the SHMR. In deriving the mass bins within each type to perform the distribution matching, we chose limits to align with the sample analyzed in \citet{Amodeo2021}.  Within each of the limits, we further split into smaller bins and assigned weights based on a normalized distribution (same procedure and weights as in \citet{Amodeo2021}), shown in the right panel of Figure~\ref{fig:msmh_relation_and_cmass}. 
Since the CMASS sample was chosen spectroscopically by stellar mass, we need to convert these stellar mass bins into corresponding halo mass bins to calculate the quantities that require halo mass information. Following the same procedure as \citet{Amodeo2021}, we use the SHMR of \citet{Kravtsov2018} to perform this conversion.
Since TNG has not been tuned to match these sorts of observed relations, the Kravtsov relation does not match the TNG SHMR (see the left panel of  Figure~\ref{fig:msmh_relation_and_cmass}). By using the Kravtsov relation as the basis of converting our stellar mass weighting bins to halo masses, we can see the effects of using the same weights calculated for a stellar mass distribution (like the CMASS sample) on a sample that has been selected for halo mass. In other words, we are using the same weights (computed using the SHMR) on halos that are selected by different mass types.

The TNG sample shown in the left panel of Figure~\ref{fig:msmh_relation_and_cmass} is halo mass selected for halos with masses $11\leq\log_{10}(\frac{M}{M_\odot})\leq14$. As can be seen in the figure, the slope of the SHMR for TNG is much steeper than the relation of \citet{Kravtsov2018}. 

\subsection{Fitting Procedure of Simulated Profiles}\label{sec:fitting}
We fit the profiles computed for the different samples (red versus total, matched versus unmatched) using the simulated data ($\vec{d}$) and errors ($\vec{\Sigma}$), given by the $\pm 1\sigma$ distribution of profiles in each radial bin shown in Figure~\ref{fig:prof}. We assume the likelihood, $\mathrm{L}(\vec{d})$, to have the form
\begin{equation}
    \ln \mathrm{L}(\vec{d}) = - \frac{1}{2} \left[ \vec{d}-\vec{\mu}(\vec{\theta}) \right]^{\rm T} \vec{\Sigma}^{-1} \left[ \vec{d}-\vec{\mu}(\vec{\theta}) \right] \,,
    \label{eq:likelihood}
\end{equation}
where $\vec{\mu}(\vec{\theta})$ is the model evaluated with parameters $\vec{\theta}$. The posterior, $\mathrm{P}(\vec{\theta})$, is written in terms of the likelihood as
\begin{equation}
  \mathrm{P}(\vec{\theta}) \propto \mathrm{L}(\vec{d}) \mathrm{Pr}(\vec{\theta}) \,,
  \label{eq:posterior}
\end{equation}
where $\mathrm{Pr}(\vec{\theta})$ are the priors on the parameters $\vec{\theta}$. We maximize the posterior probability functions using the Markov chain Monte Carlo  (MCMC) calculation package \texttt{emcee} \citep{Foreman2013} to quantify the different effects and biases of model selection (one-halo versus two-halo term GNFW model, matched versus unmatched) and sample selection (red versus total, stellar mass versus halo mass).

For the models $\vec{\mu}(\vec{\theta})$ in Equations~\ref{eq:likelihood} and~\ref{eq:posterior} we use a generalized Navarro-Frenk-White (GNFW) profile (\citet{gnfw}, see also \citet{Hernquist1990,Navarro1997}), as it is a simple parametric model that captures the shape of the density and pressure profiles \citep{Nagai2007}. 

The GNFW gas density profile has the functional form
\begin{equation}
    \label{eq:rho_gnfw}
    \begin{aligned}
        \frac{\rho(x)}{\rho_{cr}(z)} &=  f_{b} \rho_{0}(x/x_{c,k})^{\gamma_k}[1+(x/x_{c,k})^{\alpha_k}]^{-\frac{\beta_k-\gamma_k}{\alpha_k}} \, , \\
        \rho_{cr}(z) &= \frac{3H_0^2}{8\pi G}[\Omega_m(1+z)^3+\Omega_\Lambda] 
    \end{aligned}
\end{equation}
where scaled radius $x \equiv r/r_{200c}$, $f_{b}$ is the baryon fraction $\Omega_b/\Omega_m$, $\rho_{cr}(z)$ is the critical density of the universe at redshift $z$, $H_0$ is the Hubble constant, and $G$ is the gravitational constant.  

The GNFW thermal pressure profile has the functional form
\begin{equation}
    \label{eq:pth_gnfw}
    \begin{aligned}
    \frac{P(x)}{P_{200c}} &=P_{0}(x/x_{c,t})^{\gamma_t}[1+(x/x_{c,t})^{\alpha_t}]^{-\beta_t} \, , \\
    P_{200c} &= \frac{200GM_{200c}\rho_{cr}(z)f_{b}}{2r_{200c}}
    \end{aligned}
\end{equation}
where $M_{200c}$ is the mass of the sphere whose mean density is 200 times the critical density of the universe at a redshift $z$. In both Equations~\ref{eq:rho_gnfw} and~\ref{eq:pth_gnfw}, $x_c$ is the core radius, $\alpha$ is the slope of the GNFW profile at $x\sim 1$, $\beta$ is the outer slope of the profile at $x \gg 1$, and $\gamma$ is the inner slope of the profile at $x \ll 1$ \citep{gnfw}. 

For the gas density profiles (Equation~\ref{eq:rho_gnfw}) we perform fits for parameters $\rho_0$, $\alpha_k$, and $\beta_k$, keeping parameters $x_{c,k}$ and $\gamma_k$ fixed due to degeneracies, described in \citet{Battaglia2016}. The values we adopt for these fixed parameters along with the flat priors on the free parameters ($\mathrm{Pr(\vec{\theta})}$ in Equation~\ref{eq:posterior}) are shown in Table~\ref{tab:gnfw_fits}. The fits are not sensitive to choice of initial states, but we use the fitting formulas derived in \citet{Battaglia2016} as the starting points for each of the MCMC chains.

For the thermal pressure profiles (Equation~\ref{eq:pth_gnfw}) we perform fits for parameters $P_0$, $x_{c,t}$, and $\beta_t$, keeping parameters $\alpha_t$ and $\gamma_t$ fixed as described in \citet{Battaglia2012b}. Similarly to the treatment of the density fits, we use the fitting formulas derived in \citet{Battaglia2012b} as the starting points for each of the MCMC chains.

\begin{table}
    \centering
    \begin{tabular}{|c|c|}
         \hline
         \multicolumn{2}{|c|}{Parameters of GNFW Profile Fits}  \\
         \hline
         Pressure & Density \\
         \hline
         \multicolumn{2}{|c|}{Fixed} \\
         \hline
         $\alpha_t$ = 1.0 & $x_{c,k}$ = 0.5 \\
         $\gamma_t$ = -0.3 & $\gamma_k$ = -0.2 \\
        \hline
        \multicolumn{2}{|c|}{Free} \\
        \hline
        $P_0$, $x_{c,t}$, $\beta_t$, $A_{t2h}$ & $\rho_0$, $\alpha_k$, $\beta_k$, $A_{k2h}$ \\
        $P_0: (0.1,30)$ & $\log_{10}\rho_0: (1.0,6.0)$ \\
        $x_{c,t}: (0.01,1.0)$ & $\alpha_k: (0.1,6.0)$ \\
        $\beta_t: (1.0,10.0)$ & $\beta_k: (1.0,10.0)$ \\
        $A_{t2h}: (0.01,5.0)$ & $A_{k2h}: (0.01,5.0)$ \\
        \hline
    \end{tabular}
    \caption{Values for the fixed parameters and free parameters in the GNFW profile fits, where $A_{2h}$ is the amplitude of the two-halo term. Priors for the fits are given for the free parameters.}\label{tab:gnfw_fits} 
\end{table}

As in \citet{Amodeo2021}, we add in the contributions of neighboring halos with the addition of a two-halo term multiplied by an amplitude, $A_{2h}$, which is a free parameter in the fits, such that the final forms of the fitting equations are

\begin{equation}
    \begin{aligned}
    P_{GNFW1h} &= P_{1h} \, , \\ P_{GNFW} &= P_{1h}+A_{t2h}P_{2h} \label{eq:rho_fitting_eq}
    \end{aligned}
\end{equation}
where $P_{GNFW1h}$ is the model used when just taking the one-halo term into account, and $P_{GNFW}$ is the model used for taking both one-halo and two-halo terms into the fit. The fitting equations for the different models of density similarly take this form.

We run several independent MCMC chains for each sample until the Gelman-Rubin statistic \citep{Gelman1992} reaches values of $\lesssim 1.1$. We combine the independent chains and derive marginalized estimates for each parameter using the median $\pm 1\sigma$ as the error estimates. We select the model resulting in the minimum $\chi^2$ as our \textit{best} fit, shown as the fits in Figure~\ref{fig:fits3d}. 

\section{Observational Profiles}\label{sec:methods_2d}

In order to understand how the above modeling choices described in Section~\ref{sec:modeling} impact the CGM properties that we will infer from observations we need to forward model the simulated profiles from TNG into an observable space, e.g., projected SZ profiles. Here we choose to model a subsample of CMASS, as recent SZ profile measurements of this sample have been made \citep{Amodeo2021,Schaan2021}. The general outline of this procedure is to make the modeling selections within the 3D simulated TNG halos, project the computed profiles to how they would actually be observed in the 2D observing space of SZ effect measurements, and fit the profiles with the same process as described in Section~\ref{sec:fitting}.

\subsection{Modeling the Sample Selection}\label{sec:modeling_2d}
For our fiducial model, we design our simulated halo sample and fitting model to best reflect the properties of the observed CMASS halo sample.

We use the halos of a single TNG snapshot at $z=0.55$ to match the median redshift of the CMASS sample, $z=0.55$ \citep{Ahn2014}. We use \texttt{Illstack} to select for stellar masses to match the sample analyzed in \citet{Amodeo2021} shown in the right panel of Figure~\ref{fig:msmh_relation_and_cmass}, and further separate into finer mass bins within this range to weight the TNG sample to match the observed mass distribution. More specifically, the stellar mass limits of the sample are $10.71 \leq M \leq 11.72$, and the halo mass limits of the sample are $12.12 \leq M \leq 13.98$, where $M = \log_{10}(M_x/M_{\odot})$ and $M_x$ is either stellar or halo mass.
CMASS galaxies are large red galaxies \citep[LRGs;][]{Padmanabhan2007}, so we use \texttt{Illstack} to select for red halos as defined by the TNG color cut described in Section~\ref{sec:color_splits}. Lastly, we include a two-halo term to the model as described in Section~\ref{sec:fitting}. Table~\ref{tab:systematics} summarizes the labels for each modeling uncertainty.

\begin{table*}
    \centering
    \begin{tabular}{|c|c|}
         \hline
         \multicolumn{2}{|c|}{Sample Selection} \\
         \hline
         ``ms" = Stellar mass-selected & ``red" = Color-selected, $g-r > 0.6$ \\
         ``mh" = Halo mass-selected & ``tot" = No color selection \\
        \hline
        \multicolumn{2}{|c|}{Fitting Model} \\
        \hline
        ``m" = CMASS mass distribution matched & ``GNFW" = With two-halo term \\
        ``um" = Unmatched & ``GNFW1h" = No two-halo term \\
        \hline
    \end{tabular}
    \caption{Explanation of the labels used to identify the choices included for the halo sample and fitting model, see Section~\ref{sec:modeling_2d}.}\label{tab:systematics}
\end{table*}

\subsection{Generating 2D Profiles}\label{sec:projections}
The signals of the kSZ and tSZ effects are integrated along the LOS, so we need to project our simulated 3D profiles in the same way. We use the repository \texttt{Mop-c-GT}\footnote{\url{https://github.com/samodeo/Mop-c-GT}}, (Model-to-observable projection code for Galaxy Thermodynamics), introduced in \citet{Amodeo2021}. This repository inputs 3D gas density and pressure radial profiles and outputs 2D profiles of observable quantities through LOS projection. Specifically, it produces profiles of the temperature shifts in CMB signal due to the kSZ and tSZ effects, which we derive from the simulated gas density and pressure profiles, respectively. 
For the kSZ signal, we assume that the RMS of the peculiar velocities projected along the LOS is 313 km/s, as predicted by the linear theory at $z=0.55$ and adopted in \citet{Schaan2021}. This value only impacts the amplitude of the signal and is a systematic uncertainty in the velocity reconstruction, which is beyond the scope of this analysis. \texttt{Mop-c-GT} also
allows for forecasting of signals received by individual instruments with instrumental beam convolution, and an aperture photometry filter is also used, described in more detail in \citet{Amodeo2021} and \citet{Schaan2021}.
Here we convolve the profiles with a Gaussian beam of 1.4$^\prime$ at frequency 150 GHz, corresponding to the forecasted Simons Observatory (SO) experimental setup described in \citet{SO2019}.
We are using a method to create stacked profiles of multiple hundreds of halos. The halos could have differing individual orientations, but by stacking together we are able to assume a spherically symmetric profile. Thus, changing the direction of projection should not yield different results (see \citet{Battaglia2012a}). To further demonstrate that additional contributions along the LOS are completely subdominant, we performed a test extending the projected profiles out to $\sim$50 Mpc to see whether the contributions to the profile from larger radii affect the results of the projection process. We find that the difference in the projected SZ profiles extending to 10 Mpc and extending to 50 Mpc is completely negligible, $\ll 0.1\%$, which highlights the benefit of using the aperture photometry filter.

Similar analyses and profiles can be made for other CMB experiments, such as CMB-S4, and the subsequent error bars will change with the different sensitivities, frequency channels, and beams \citep[e.g., see section 1.4.2.1 in the CMB-S4 Science Case and Reference Design document,][]{CMB-S4-SRD}.

\subsection{Fitting Procedures for Observational Profiles}\label{sec:fitting_2d}
We compute the 2D observational profile using the parameters $\vec{\theta}$ from the best fit of the 3D theoretical profile using \texttt{Mop-c-GT}, with the same GNFW models $\vec{\mu}(\vec{\theta})$ as defined in Section~\ref{sec:fitting}. We use this profile as the data, $\vec{d}$, to calculate the likelihood of the same form as Equation~\ref{eq:likelihood}. However, instead of using the distribution of the 3D profiles as the errors $\vec{\Sigma}$ we use the covariance matrix of forecasted errors for the SO experiment of \citet{Battaglia2017}, in which the authors use a semi-analytical foreground model that includes contributions from the cosmic infrared background, primary CMB fluctuations, extragalactic radio emission, and galactic cirrus. 

We fit for the same parameters ($\vec{\theta}$) for each kind of profile and use the same priors, $\mathrm{Pr({\vec{\theta}})}$, previously described in Section~\ref{sec:fitting} and shown in Table~\ref{tab:gnfw_fits}.

\section{Results}

\subsection{3D Simulations}\label{sec:results_3d}
Here we show the results of fitting the 3D simulated profiles, focusing on the different modeling choices and any trends in mass. The parameters of the fits for each sample and model are shown in Table~\ref{tab:fits_3d}.

\subsubsection{Modeling Choices}
\begin{figure*}
    \centering
    \includegraphics[scale=0.6]{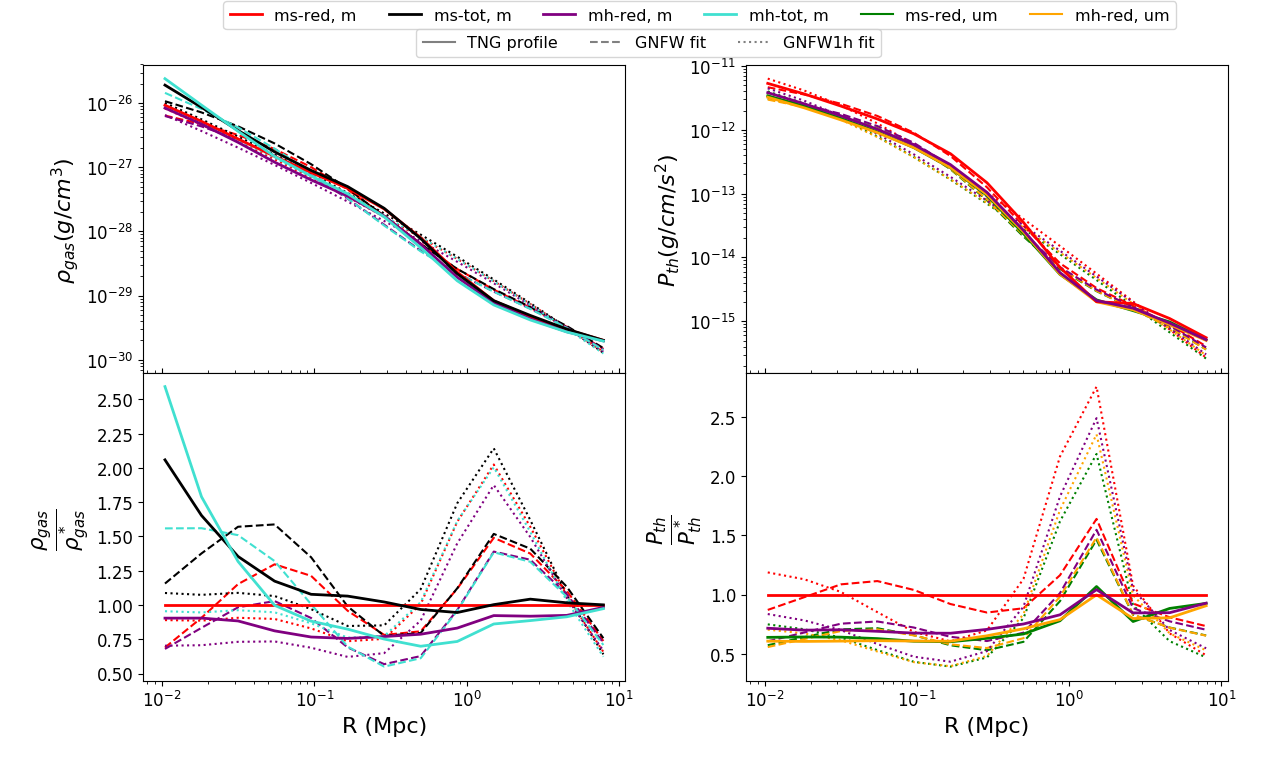}
    \vspace{-0.3cm}
    \caption{\label{fig:fits3d} Simulated profiles of density (left) and pressure (right) showing the populations fit with different mass selections, color selections, distribution-matching options, and fitting models. The solid lines are the TNG profiles for the population specified by the legend, the dashed lines are the best fit from the GNFW model, and the dotted lines are the best fit from the GNFW1h model. In the bottom panels, the profiles and fits are normalized by a fiducial TNG profile for sample ms-red, m (solid red curve) denoted as $\rho_{gas}^*$ and $P_{th}^*$. This figure shows that selections by color and mass type (seen in the left panels) and the distribution matching (seen in the right panels) have varying levels of significance on the profiles. Furthermore, this figure shows the inclusion of a two-halo term to the fitting model provides a closer match to the profile than a model that only includes the one-halo term.}
\end{figure*}

As a qualitative example to demonstrate the effects of different modeling procedures, we show fits of the 3D profiles in Figure~\ref{fig:fits3d}. The profiles computed from TNG are shown by solid lines, while the best GNFW and GNFW1h fits are shown by dashed and dotted lines, respectively. To be able to see differences more clearly, all of the profiles and fits have been normalized by the fiducial TNG profile for the ms-red, m sample, denoted as $\rho_{gas}^*$ and $P_{th}^*$ in the bottom panels of the figure. We explore different modeling choices in each column, though the trends are similar for both pressure and density. Both columns show profiles of differing mass selections; additionally, in the left column we show profiles differing in color selection, and in the right column we show profiles differing in distribution-matching options.

First, the left column of Figure~\ref{fig:fits3d} shows the differences in sample selection, including mass type and color, for the distribution-matched density profiles. While the inner regions of the profiles for the ms-tot and mh-tot samples have higher values than the other profiles, it can be seen that for the rest of the profiles, the ms samples tend to have higher values than the mh samples. Beyond the inner part of the profiles, the color selection does not appear to have a significant effect: for the mh samples the red color-selected sample has higher values than the corresponding tot sample (compare solid purple curve with solid turquoise curve), and within the ms samples the profiles of the red and tot samples are nearly identical (compare the solid red and solid black curves). 

Next, the right column of Figure~\ref{fig:fits3d} shows the differences in distribution-matching options and mass selection of the pressure profiles. A separation between the matched and unmatched profiles can be seen in both panels, with the unmatched profiles having lower values ($\lesssim 0.7$ times the fiducial profile $P_{th}^*$, which is matched) than their matched counterparts. This is expected since the matched and unmatched samples are compiled from different halo distributions (Figure~\ref{fig:msmh_relation_and_cmass}). Mass distribution matching is expected to have a larger effect on the pressure profiles than the density profiles, due to the pressure's dependence on $M^{5/3}$. Differences are seen among the density profiles as well, although to a lesser extent. Similarly to the density profiles of the left column, within each of the weighting groupings the profiles of the ms samples tend to have higher values than those of the mh samples.

Lastly, Figure~\ref{fig:fits3d} shows how the fits vary for the different fitting models, GNFW versus GNFW1h. An obvious feature in both bottom panels of the figure is the spike in values around 1 Mpc, although more significant for the GNFW1h model (dotted) and more significant for pressure than density. This indicates the fits' inabilities to accurately reflect the shape of the outer profiles, and in this case the best GNFW1h models return higher values than the actual profiles at these radii by factors of 2 or 3. It can be seen by eye that the GNFW model provides better fits to the profiles, as the spikes are less significant and they more closely align with the actual profiles at other radii. 

\subsubsection{Mass Dependence}\label{sec:results_mass_dependence}
\begin{figure}
    \centering
    \includegraphics[scale=0.45]{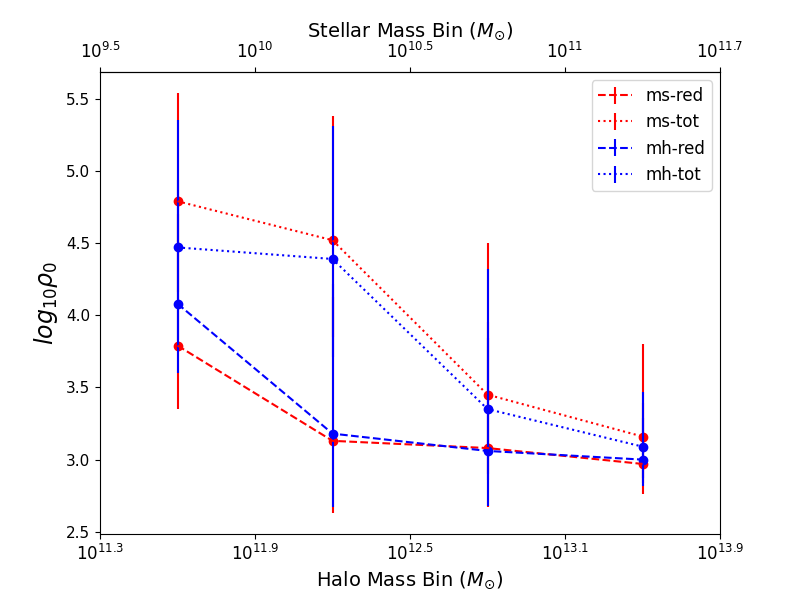}
    
    \includegraphics[scale=0.45]{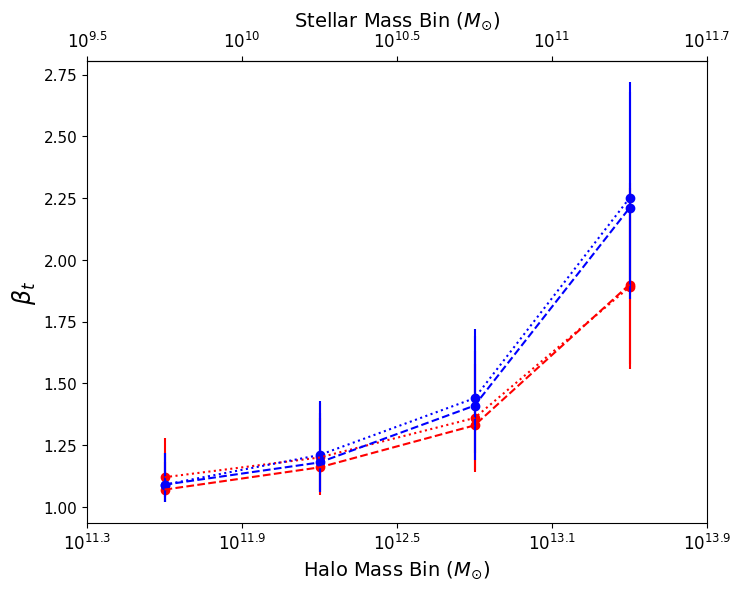}

    \caption{Mass dependence of marginalized parameters for each sample fit by the GNFW model (top) and GNFW1h model (bottom). The point in each bin shows the median of the MCMC chains with error bars $\pm 1\sigma$ showing the distribution within the bin. The red and blue lines (showing different mass selections) have been slightly offset in each mass bin to more clearly show the different error bars, along with different cap sizes to more clearly show the error bars of the red versus tot samples.}
    \label{fig:fits_3d_mass}
\end{figure}

As described in Section~\ref{sec:mass_splits}, we split the simulated halos into mass bins to observe any trends in our models. In Figure~\ref{fig:fits_3d_mass} we show the marginalized estimates for parameters $\rho_0$ and $\beta_t$, resulting from the MCMC chains.
The points in each mass bin show the median of all the samples with the error bars showing the 1$\sigma$ range. The top and bottom axes show the stellar mass and halo mass bins used to select the halos, respectively.
We note that the top and bottom axes are not related by an SHMR but they just show the mass bins further divided, therefore one should not compare the ms versus mh (red versus blue) values.

Some of the parameters have flat trends with mass, but others have relations like those shown in Figure~\ref{fig:fits_3d_mass}. The top panel shows that the normalization factor for the GNFW density profile has a negative trend scaling with mass, and the bottom panel shows that $\beta_t$ for the GNFW1h pressure profile has a positive trend with mass. The CMASS sample spans a large range in masses and such a range is probably not unique to CMASS. Figure~\ref{fig:fits_3d_mass} shows that some parameters show mass dependencies while others do not. Capturing such mass trends is important when modeling observable samples for cross correlations, and is a motivating reason for why we looked into matching the sample to the observed mass distribution.

\begin{table*}
    \centering
    \resizebox{\textwidth}{!}{\begin{tabular}{ccccc|ccccc}
        \toprule
        \multicolumn{10}{c}{Parameters from 3d Fits} \\
        \midrule
        \multicolumn{10}{c}{Model GNFW-matched} \\
        \multicolumn{5}{c}{Density} & \multicolumn{5}{c}{Pressure} \\
        \hline
        & mstar-tot & mstar-red & mh-tot & mh-red & & mstar-tot & mstar-red & mh-tot & mh-red \\
        \hline
        $\log_{10}\rho_0$ & 3.67, $3.04_{-0.28}^{+0.60}$ & 3.28, $2.92_{-0.21}^{+0.41}$ & 4.34, $3.22_{-0.42}^{+0.84}$ & 3.38, $2.88_{-0.23}^{+0.46}$ & $P_0$ & 4.18, $8.86_{-4.17}^{+8.25}$ & 4.03, $8.66_{-4.02}^{+7.89}$ & 2.74, $6.94_{-3.62}^{+8.50}$ & 2.86, $7.45_{-3.96}^{+8.77}$ \\ 
        $\alpha_k$ & 0.64, $1.49_{-0.79}^{+1.66}$ & 0.80, $1.65_{-0.84}^{+1.62}$ & 0.43, $1.10_{-0.56}^{+1.19}$ & 0.68, $1.67_{-0.88}^{+1.78}$ & $x_{c,t}$ & 0.97, $0.59_{-0.30}^{+0.27}$ & 1.00, $0.60_{-0.30}^{+0.27}$ & 0.98, $0.51_{-0.28}^{+0.31}$ & 0.93, $0.53_{-0.29}^{+0.30}$ \\ 
        $\beta_k$ & 3.37, $4.31_{-1.06}^{+2.01}$ & 3.20, $4.10_{-1.01}^{+1.99}$ & 3.44, $4.76_{-1.31}^{+2.42}$ & 3.30, $4.90_{-1.55}^{+2.67}$ & $\beta_t$ & 6.09, $5.46_{-2.07}^{+2.42}$ & 6.14, $5.43_{-2.00}^{+2.36}$ & 6.31, $5.65_{-2.40}^{+2.65}$ & 5.80, $5.67_{-2.31}^{+2.59}$ \\
        $A_{k2h}$ & 1.40, $1.60_{-0.36}^{+0.43}$ & 1.37, $1.57_{-0.36}^{+0.43}$ & 1.31, $1.50_{-0.34}^{+0.43}$ & 1.34, $1.53_{-0.34}^{+0.41}$ & $A_{t2h}$ & 0.57, $0.95_{-0.47}^{+0.82}$ & 0.53, $0.87_{-0.42}^{+0.72}$ & 0.51, $0.98_{-0.52}^{+1.01}$ & 0.51, $0.91_{-0.46}^{+0.82}$ \\
        \midrule
        \multicolumn{10}{c}{Model GNFW-unmatched} \\
        \multicolumn{5}{c}{Density} & \multicolumn{5}{c}{Pressure} \\
        \hline
        & mstar-tot & mstar-red & mh-tot & mh-red & & mstar-tot & mstar-red & mh-tot & mh-red \\
        \hline
        $\log_{10}\rho_0$ & 4.11, $3.25_{-0.42}^{+0.89}$ & 3.38, $2.98_{-0.26}^{+0.59}$ & 5.68, $3.62_{-0.58}^{+1.04}$ & 3.48, $2.91_{-0.26}^{+0.57}$ & $P_0$ & 3.95, $9.31_{-4.53}^{+8.70}$ & 4.91, $10.70_{-5.05}^{+8.88}$ & 3.84, $10.16_{-5.25}^{+9.56}$ & 4.47, $10.85_{-5.56}^{+9.59}$ \\ 
        $\alpha_k$ & 0.49, $1.02_{-0.51}^{+1.09}$ & 0.68, $1.24_{-0.62}^{+1.16}$ & 0.30, $0.83_{-0.38}^{+0.77}$ & 0.61, $1.42_{-0.75}^{+1.48}$ & $x_{c,t}$ & 0.98, $0.59_{-0.31}^{+0.28}$ & 0.98, $0.62_{-0.31}^{+0.26}$ & 0.98, $0.59_{-0.30}^{+0.28}$ & 0.95, $0.60_{-0.30}^{+0.27}$ \\ 
        $\beta_k$ & 3.30, $3.85_{-0.79}^{+1.32}$ & 2.91, $3.41_{-0.71}^{+1.29}$ & 3.54, $4.31_{-0.93}^{+1.54}$ & 3.03, $4.01_{-1.10}^{+2.20}$ & $\beta_t$ & 5.00, $4.75_{-1.80}^{+2.42}$ & 4.65, $4.48_{-1.63}^{+2.26}$ & 4.61, $4.80_{-2.02}^{+2.72}$ & 4.55, $4.79_{-1.88}^{+2.61}$ \\
        $A_{k2h}$ & 1.31, $1.47_{-0.33}^{+0.40}$ & 1.28, $1.47_{-0.34}^{+0.41}$ & 1.20, $1.36_{-0.31}^{+0.38}$ & 1.30, $1.50_{-0.34}^{+0.41}$ & $A_{t2h}$ & 0.49, $0.84_{-0.41}^{+0.71}$ & 0.47, $0.80_{-0.39}^{+0.64}$ & 0.44, $0.86_{-0.46}^{+0.86}$ & 0.47, $0.91_{-0.46}^{+0.82}$ \\
        \midrule
        \multicolumn{10}{c}{Model GNFW1h-matched} \\
        \multicolumn{5}{c}{Density} & \multicolumn{5}{c}{Pressure} \\
        \hline
        & mstar-tot & mstar-red & mh-tot & mh-red & & mstar-tot & mstar-red & mh-tot & mh-red \\
        \hline
        $\log_{10}\rho_0$ & 6.00, $4.68_{-1.01}^{+0.89}$ & 6.00, $4.53_{-0.96}^{+0.95}$ & 6.00, $4.76_{-1.00}^{+0.84}$ & 6.00, $4.45_{-1.00}^{+1.00}$ & $P_0$ & 18.33, $10.02_{-5.93}^{+10.94}$ & 17.37, $10.24_{-6.01}^{+10.81}$ & 13.26, $6.51_{-4.07}^{+10.91}$ & 13.80, $7.32_{-4.71}^{+11.28}$ \\ 
        $\alpha_k$ & 0.17, $0.25_{-0.06}^{+0.14}$ & 0.16, $0.25_{-0.07}^{+0.14}$ & 0.17, $0.24_{-0.05}^{+0.12}$ & 0.16, $0.25_{-0.07}^{+0.15}$ & $x_{c,t}$ & 0.06, $0.14_{-0.08}^{+0.21}$ & 0.06, $0.14_{-0.07}^{+0.19}$ & 0.05, $0.15_{-0.10}^{+0.31}$ & 0.05, $0.14_{-0.09}^{+0.26}$ \\ 
        $\beta_k$ & 2.06, $1.95_{-0.18}^{+0.16}$ & 1.99, $1.90_{-0.16}^{+0.15}$ & 2.03, $1.94_{-0.17}^{+0.16}$ & 1.92, $1.83_{-0.17}^{+0.16}$ & $\beta_t$ & 1.53, $1.80_{-0.33}^{+0.44}$ & 1.54, $1.81_{-0.33}^{+0.43}$ & 1.42, $1.73_{-0.35}^{+0.49}$ & 1.41, $1.72_{-0.35}^{+0.47}$ \\
        \midrule
        \multicolumn{10}{c}{Model GNFW1h-unmatched} \\
        \multicolumn{5}{c}{Density} & \multicolumn{5}{c}{Pressure} \\
        \hline
        & mstar-tot & mstar-red & mh-tot & mh-red & & mstar-tot & mstar-red & mh-tot & mh-red \\
        \hline
        $\log_{10}\rho_0$ & 6.00, $4.80_{-1.01}^{+0.82}$ & 6.00, $4.55_{-1.01}^{+0.96}$ & 6.00, $4.92_{-0.99}^{+0.75}$ & 6.00, $4.40_{-1.04}^{+1.04}$ & $P_0$ & 19.54, $9.58_{-5.70}^{+11.05}$ & 19.23, $10.53_{-5.99}^{+10.73}$ & 16.64, $7.79_{-4.59}^{+11.06}$ & 16.50, $8.88_{-5.27}^{+11.09}$ \\ 
        $\alpha_k$ & 0.16, $0.23_{-0.05}^{+0.11}$ & 0.16, $0.24_{-0.06}^{+0.14}$ & 0.16, $0.22_{-0.04}^{+0.10}$ & 0.15, $0.24_{-0.07}^{+0.16}$ & $x_{c,t}$ & 0.06, $0.16_{-0.09}^{+0.23}$ & 0.08, $0.19_{-0.10}^{+0.25}$ & 0.07, $0.22_{-0.13}^{+0.34}$ & 0.08, $0.21_{-0.13}^{+0.31}$ \\ 
        $\beta_k$ & 1.97, $1.87_{-0.17}^{+0.16}$ & 1.88, $1.78_{-0.16}^{+0.15}$ & 1.94, $1.84_{-0.17}^{+0.15}$ & 1.84, $1.73_{-0.17}^{+0.16}$ & $\beta_t$ & 1.39, $1.64_{-0.30}^{+0.38}$ & 1.42, $1.67_{-0.31}^{+0.38}$ & 1.29, $1.57_{-0.30}^{+0.38}$ & 1.39, $1.69_{-0.33}^{+0.41}$ \\
        \bottomrule
    \end{tabular}}
    \caption{Parameters from 3D fits for each sample and model. The best values (minimum $\chi^2$) for each fit are listed first, followed by the marginalized parameter listed as the median $\pm 1\sigma$.} 
    \label{tab:fits_3d} 
\end{table*}

\subsection{Modeled Observations}\label{sec:results_2d}
\begin{figure*}
	\centering
	\includegraphics[scale=0.68]{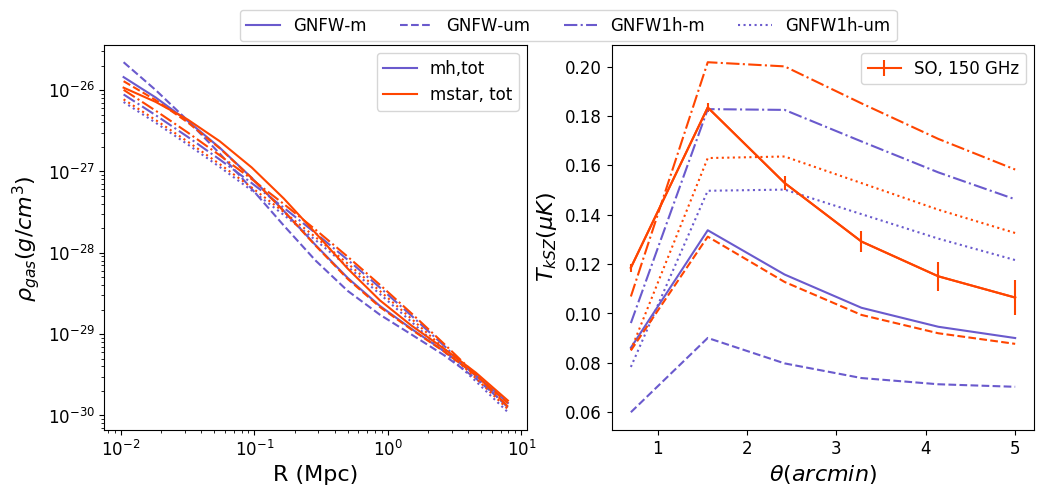}

	\includegraphics[scale=0.68]{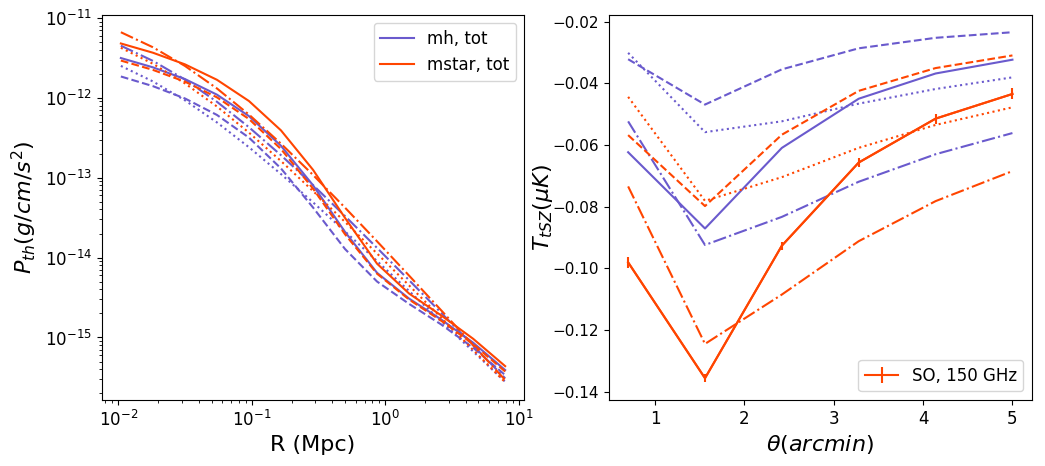}
	\caption{\label{fig:projections} Projection process of \texttt{Mop-c-GT}. Input 3D theoretical profiles (in this case, from TNG) and output the tSZ and kSZ observable temperature shifts as a function of radius in arcminutes. The different colors correspond to different mass selections, and different line styles correspond to different fitting models and inclusion of distribution matching. The error bars correspond to the forecasted SO error bars at 150 GHz, discussed further in Section~\ref{sec:fitting_2d}. These profiles show the effects of 1. selecting by different mass types, with the ms samples having higher values and amplitudes than the mh samples, 2. mass-distribution matching, with more significant differences for pressure but still relevant for density, and 3. inclusion of a two-halo term, which results in lower amplitudes of the observed profiles as the 1h-only model overpredicts the values.}
\end{figure*}

Here we show the results of projecting and fitting the 2D simulated profiles. Figure~\ref{fig:projections} shows the projection process using \texttt{Mop-c-GT}; the left columns show the fits of the 3D simulated profiles from TNG for different halo selections and fitting models, which are input into the projection code, and the right columns show these different populations projected into tSZ and kSZ temperature shifts described in Section~\ref{sec:projections}. The different line styles correspond to the different kind of fit, as specified in the legend of the figure, and all of the fits are selected by minimum $\chi^2$. The error bars are the forecasted errors for the SO experiment.

The top panels of Figure~\ref{fig:projections} show fits for the density profiles. While not as prominent as the differences in pressure, the inclusion of mass-distribution matching does have an effect on the observed density. In general, a profile with higher 3D values will have higher measurable signals, which is what we see in this figure. The GNFW and GNFW1h-matched models have higher values than the unmatched fits in 3D, and thus have higher kSZ signals. Within each of the fitting models the stellar mass-selected sample has higher kSZ signals than the halo mass-selected sample, which is also seen in the 3D profiles. This is the result of SHMR of IllustrisTNG being lower than the \citet{Kravtsov2018} relation, so at a fixed stellar mass TNG will select a higher halo mass. Furthermore, the GNFW1h models overpredict the values of the profiles at high radii (seen previously in Figure~\ref{fig:fits3d}), so the amplitudes of the profiles only taking into account the one-halo term are higher than the models including a two-halo term. 

The lower panels of Figure~\ref{fig:projections} shows fits for the pressure profiles. As shown previously in Figure~\ref{fig:fits3d}, the inclusion of mass-distribution matching has a significant effect on the 3D pressure profiles due to its dependence on $M^{5/3}$. We can see this in more detail in both bottom panels, along with similar trends to the top panel density profiles with the matched models and the stellar mass-selected sample resulting in higher values and amplitudes.
It is clear from Figure~\ref{fig:projections} that the modeling choices for the cross-correlation signal are important simply by comparing the differences in the observed profiles to the size of the forecasted errors.
 
\begin{table*}[!ht]
    \centering
    \resizebox{\textwidth}{!}{\begin{tabular}{ccccc|ccccc}
        \toprule
        \multicolumn{10}{c}{Parameters from 2d Fits} \\
        \midrule
        \multicolumn{10}{c}{Model GNFW-matched} \\
        \multicolumn{10}{c}{(Input Parameters for Fits Below)} \\
        \multicolumn{5}{c}{Density} & \multicolumn{5}{c}{Pressure} \\
        \hline
        & ms-tot & ms-red & mh-tot & mh-red & & ms-tot & ms-red & mh-tot & mh-red \\
        \hline
        $\log_{10}\rho_0$ & $3.67_{-0.05}^{+0.05}$ & $3.28_{-0.04}^{+0.04}$ & $4.34_{-0.10}^{+0.10}$ & $3.39_{-0.06}^{+0.07}$ & $P_0$ & $4.20_{-0.23}^{+0.25}$ & $4.05_{-0.22}^{+0.24}$ & $2.78_{-0.25}^{+0.28}$ & $2.89_{-0.22}^{+0.25}$ \\ 
        $\beta_k$ & $3.37_{-0.11}^{+0.11}$ & $3.20_{-0.11}^{+0.11}$ & $3.45_{-0.16}^{+0.16}$ & $3.30_{-0.16}^{+0.16}$ & $\beta_t$ & $6.10_{-0.15}^{+0.16}$ & $6.16_{-0.15}^{+0.16}$ & $6.36_{-0.25}^{+0.27}$ & $5.83_{-0.20}^{+0.21}$ \\
        $A_{k2h}$ & $1.41_{-0.23}^{+0.23}$ & $1.37_{-0.23}^{+0.23}$ & $1.32_{-0.23}^{+0.23}$ & $1.35_{-0.23}^{+0.23}$ & $A_{t2h}$ & $0.57_{-0.06}^{+0.06}$ & $0.53_{-0.06}^{+0.06}$ & $0.52_{-0.06}^{+0.06}$ & $0.51_{-0.06}^{+0.06}$ \\
        \midrule
        \multicolumn{10}{c}{Model GNFW-unmatched} \\
        \multicolumn{5}{c}{Density} & \multicolumn{5}{c}{Pressure} \\
        \hline
        & ms-tot & ms-red & mh-tot & mh-red & & ms-tot & ms-red & mh-tot & mh-red \\
        \hline
        $\log_{10}\rho_0$ & $3.69_{-0.05}^{+0.05}$, $0.3\sigma$ & $3.29_{-0.04}^{+0.04}$, $0.2\sigma$  & $4.42_{-0.10}^{+0.10}$, $0.6\sigma$ & $3.41_{-0.07}^{+0.07}$, $0.3\sigma$ & $P_0$ & $6.80_{-0.37}^{+0.40}$, $5.9\sigma$ & $5.98_{-0.32}^{+0.34}$, $4.9\sigma$ & $7.52_{-0.66}^{+0.76}$, $6.7\sigma$ & $4.86_{-0.37}^{+0.41}$, $4.5\sigma$ \\
        $\beta_k$ & $3.12_{-0.10}^{+0.10}$, $1.7\sigma$ & $2.95_{-0.10}^{+0.10}$, $1.7\sigma$ & $3.13_{-0.14}^{+0.14}$, $1.5\sigma$ & $3.03_{-0.14}^{+0.15}$, $1.2\sigma$ & $\beta_t$ & $4.88_{-0.11}^{+0.12}$, $6.0\sigma$ & $5.00_{-0.12}^{+0.12}$, $5.7\sigma$ & $4.60_{-0.17}^{+0.18}$, $5.3\sigma$ & $4.66_{-0.15}^{+0.16}$, $4.4\sigma$ \\
        $A_{k2h}$ & $1.38_{-0.23}^{+0.23}$, $0.1\sigma$ & $1.34_{-0.23}^{+0.23}$, $0.1\sigma$ & $1.29_{-0.24}^{+0.23}$, $0.1\sigma$ & $1.33_{-0.24}^{+0.23}$, $0.1\sigma$ & $A_{t2h}$ & $0.55_{-0.06}^{+0.06}$, $0.2\sigma$ & $0.51_{-0.06}^{+0.06}$, $0.2\sigma$ & $0.49_{-0.06}^{+0.06}$, $0.2\sigma$ & $0.49_{-0.06}^{+0.06}$, $0.2\sigma$ \\
        \hline
        \multicolumn{10}{c}{Model GNFW1h-matched} \\
        \multicolumn{5}{c}{Density} & \multicolumn{5}{c}{Pressure} \\
        \hline
        & ms-tot & ms-red & mh-tot & mh-red & & ms-tot & ms-red & mh-tot & mh-red \\
        \hline
        $\log_{10}\rho_0$ & $3.47_{-0.03}^{+0.03}$, $3.4\sigma$ & $3.12_{-0.02}^{+0.02}$, $3.6\sigma$ & $3.94_{-0.05}^{+0.05}$, $3.6\sigma$ & $3.13_{-0.04}^{+0.04}$, $3.1\sigma$ & $P_0$ & $3.11_{-0.13}^{+0.13}$, $3.8\sigma$ & $3.08_{-0.13}^{+0.13}$, $3.5\sigma$ & $1.79_{-0.11}^{+0.12}$, $3.1\sigma$ & $1.98_{-0.11}^{+0.12}$, $3.2\sigma$ \\ 
        $\beta_k$ & $2.86_{-0.06}^{+0.06}$, $4.1\sigma$ & $2.71_{-0.06}^{+0.06}$, $3.9\sigma$ & $2.79_{-0.08}^{+0.08}$, $3.6\sigma$ & $2.62_{-0.08}^{+0.08}$, $3.8\sigma$ & $\beta_t$ & $5.24_{-0.10}^{+0.10}$, $4.5\sigma$ & $5.35_{-0.10}^{+0.10}$, $4.2\sigma$ & $5.07_{-0.15}^{+0.15}$, $4.0\sigma$ & $4.79_{-0.12}^{+0.12}$, $4.2\sigma$ \\
        \hline
        \multicolumn{10}{c}{Model GNFW1h-unmatched} \\
        \multicolumn{5}{c}{Density} & \multicolumn{5}{c}{Pressure} \\
        \hline
        & ms-tot & ms-red & mh-tot & mh-red & & ms-tot & ms-red & mh-tot & mh-red \\
        \hline
        $\log_{10}\rho_0$ & $3.49_{-0.03}^{+0.03}$, $3.1\sigma$ & $3.14_{-0.02}^{+0.02}$, $3.1\sigma$ & $4.03_{-0.05}^{+0.05}$, $2.8\sigma$ & $3.16_{-0.04}^{+0.04}$, $2.7\sigma$ & $P_0$ & $5.12_{-0.20}^{+0.21}$, $2.9\sigma$ & $4.62_{-0.18}^{+0.19}$, $2.0\sigma$ & $4.92_{-0.30}^{+0.33}$, $5.3\sigma$ & $3.39_{-0.18}^{+0.19}$, $1.7\sigma$ \\ 
        $\beta_k$ & $2.67_{-0.05}^{+0.05}$, $5.8\sigma$ & $2.52_{-0.05}^{+0.05}$, $5.6\sigma$ & $2.56_{-0.07}^{+0.07}$, $5.0\sigma$ & $2.43_{-0.07}^{+0.07}$, $5.0\sigma$ & $\beta_t$ & $4.26_{-0.07}^{+0.07}$, $10.5\sigma$ & $4.41_{-0.07}^{+0.07}$, $9.9\sigma$ & $3.77_{-0.10}^{+0.10}$, $8.9\sigma$ & $3.90_{-0.09}^{+0.09}$, $8.4\sigma$ \\
        \bottomrule
    \end{tabular}}
    \caption{Parameters from 2D fits for each sample and model, with GNFW-\textit{m} as the fiducial model used as the data, $\vec d$, in the likelihood. We list the marginalized parameter for each fit as the median $\pm 1\sigma$, followed by the number of $\sigma$ away from the true value (of the GNFW-\textit{m} model). This table shows fits with an additional parameter fixed, $x_{c,t}$ for pressure and $\alpha_k$ for density, described further in Section~\ref{sec:results_2d}.} 
    \label{tab:fits_2d} 
 \end{table*}

\subsection{Testing the Number of Free Parameters}\label{sec:free-vs-fixed}
\begin{figure*}[t]
	\centering
	\includegraphics[scale=0.4]{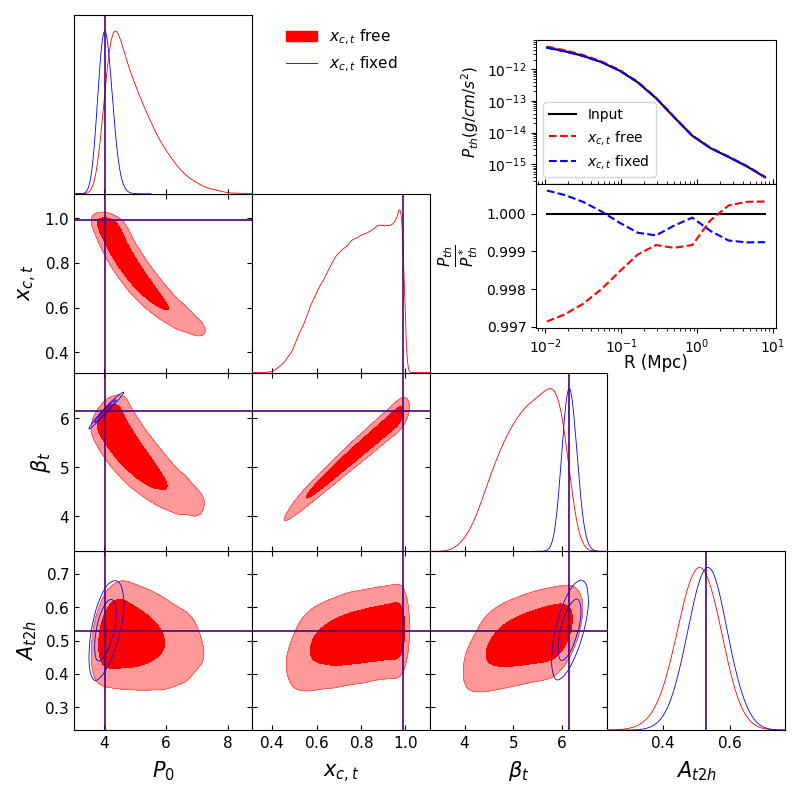}
	\includegraphics[scale=0.4]{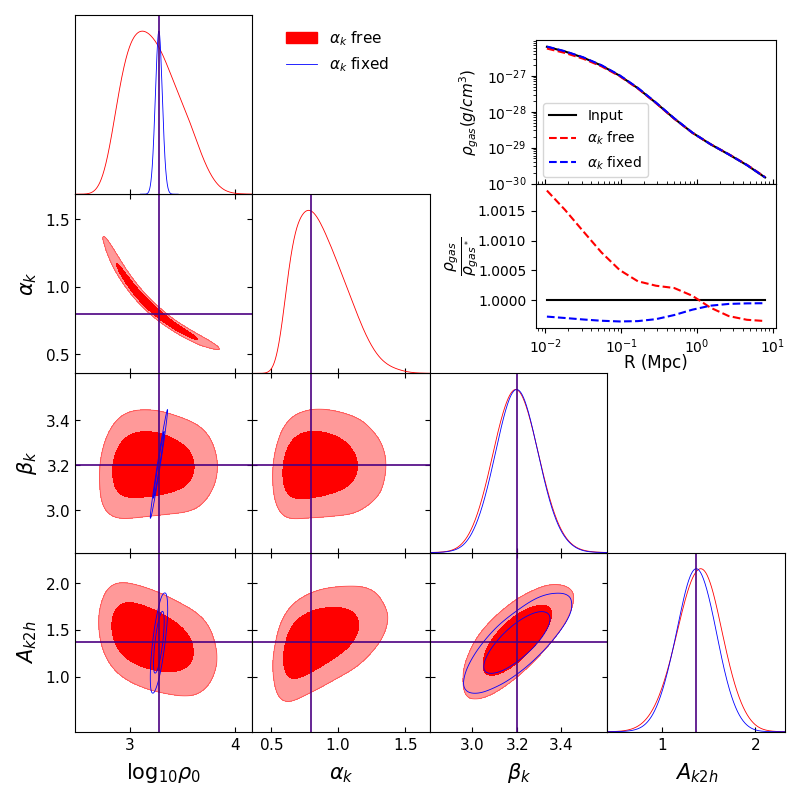}
	\vspace{-0.3cm}
	\caption{\label{fig:fixedparams} Contours for fits of projections for tSZ (left) and kSZ (right) for the GNFW-\textit{m} model of the ms-red samples. The contours for fits with all GNFW parameters free are shown in red, and the contours for fits fixing an additional parameter are shown in blue. The vertical lines show the input parameter, the best 3D fit from the corresponding sample. The small panels show the 3D profiles computed using the corresponding values for each of the different contours, using a GNFW model, and the same profiles normalized by the TNG profile to more clearly see the differences. This figure shows that the fits are not sensitive to information of the inner profiles, resulting in degeneracies, more significant for pressure. Fixing an additional parameter allows for better constraints on the remaining free parameters.}
\end{figure*}

Certain parameters of the GNFW profile are degenerate, as discussed in Section~\ref{sec:fitting}. We test how changing the number of free parameters in the fits of the observed profiles affects the results by fixing another parameter for each type of profile, shown in Figure~\ref{fig:fixedparams}. For density we fix $\alpha_k$ and for pressure we fix $x_{c,t}$, as each of these parameters have degeneracies with $\log_{10}\rho_0$ and $\beta_t$, respectively, as seen by the contours in this figure.

Figure~\ref{fig:fixedparams} shows an example of contours of fits for the GNFW-\textit{m} model of the ms-red sample using a different number of free parameters. We find that some profile information is lost in the projection process, resulting in the input 3D parameters not being recovered as well in the 2D fit with all parameters free. The solid lines show the values of the best 3D fit parameters, which serve as inputs for the 2D fits. The red contours show the fits with all of the GNFW parameters free (see Table~\ref{tab:gnfw_fits}), and the blue contours show fits holding one extra parameter constant. The small panels show the corresponding values calculated as a GNFW profile to show the differences in the 3D profiles. It can clearly be seen by the contours that fixing an extra parameter in the fits results in better alignment with the true values, as the peaks of the blue contours closely match the solid lines. Similarly, the corresponding 3D profile computed using the parameters from the blue contours more closely aligns with the 3D profile computed using the input parameters than the red.

This result indicates that with this experiment we cannot resolve the inner/intermediate parts of the profile for neither density or pressure. Perhaps with a new experiment, such as CMB-HD \citep{CMBHD} we would be able to resolve the profile further into the interior at the redshift of the CMASS sample. Furthermore, if we were to choose a different sample to model at lower redshift, we expect that our ability to resolve the inner parts of these profile would improve. Hereafter, we show the fits fixing an additional parameter to be able to constrain whether differences are due to the number of free parameters or the modeling choices.

\subsection{Biases from Modeling Choices}\label{sec:biases}

\begin{figure*}
	\centering
	\includegraphics[scale=0.5]{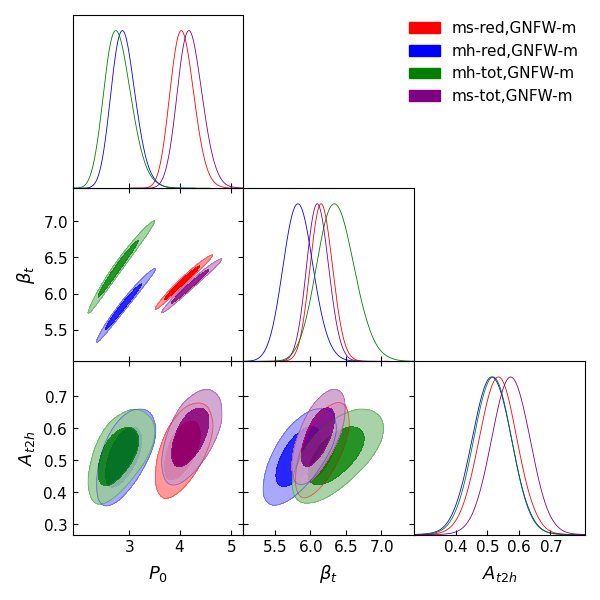}
	\includegraphics[scale=0.5]{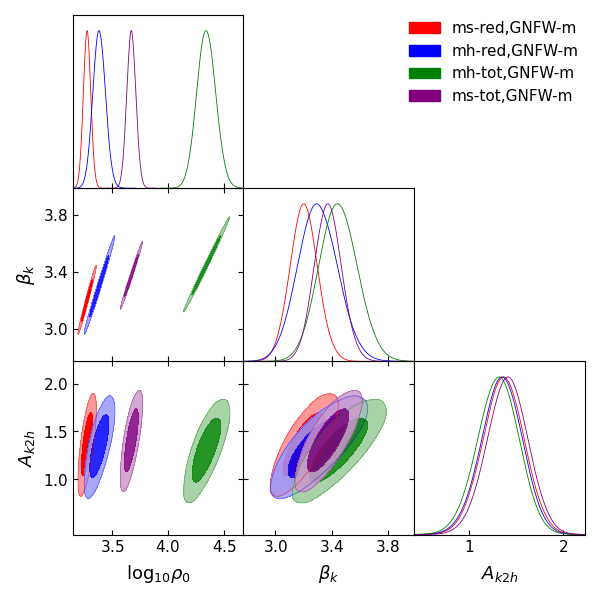}
	\vspace{-0.3cm}
	\caption{\label{fig:corner2d_samples} Comparison of parameter spaces for fits of the 2D projections of different samples, all for the GNFW-\textit{m} model. An additional parameter is fixed, $x_{c,t}$ for pressure (left) and $\alpha_k$ for density (right), as discussed in Section~\ref{sec:free-vs-fixed}. This figure shows the differences among the fits of different samples (ms versus mh, and red versus tot), highlighting the importance of these modeling choices.}
\end{figure*}

In Figure~\ref{fig:corner2d_samples} we show contours for pressure and density of different samples, all for the GNFW-\textit{m} model. These fits use different input data ($\vec{d}$) for the 2D observable profiles corresponding to each color and mass selection to show how the shapes of the profiles differ among differently-selected samples. We note the differences in two-halo term amplitudes in this work and \citet{Amodeo2021} are due to the TNG simulations having a larger two-halo term amplitude compared to the observations presented in \citet{Schaan2021}.

For pressure, there is a clear separation among the $P_0$ values favored by the halo- and stellar-mass-selected samples, with the ms samples returning higher $P_0$ values than the mh samples. The contours show more of a spread in the remaining parameters, with more significant differences in $\beta_t$ than $A_{t2h}$. For density, the normalization parameter $\log_{10}\rho_0$ shows the most significant spread among samples, with the mh-tot sample favoring the highest values and ms-red sample favoring the lowest values. For $\beta_k$ we see a spread similar to $\beta_t$ for pressure, and for $A_{k2h}$ the fits for all of the samples return nearly the same values. The differences among the contours for these samples show again that the modeling choices yield different results.

\begin{figure*}
	\centering
	\includegraphics[scale=0.5]{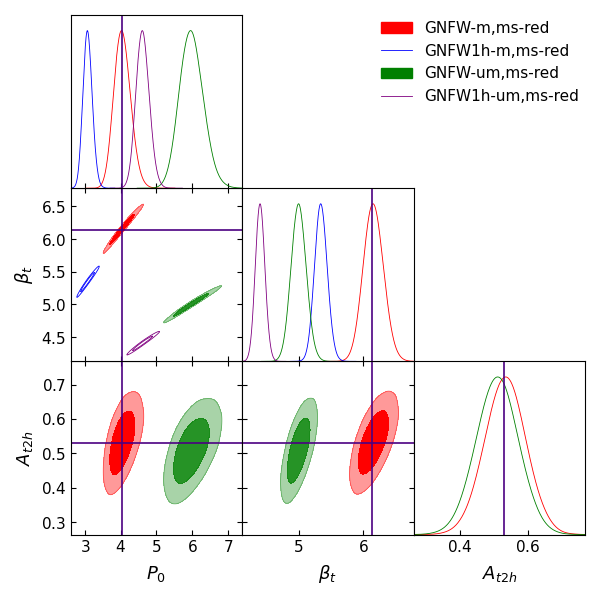}
	\includegraphics[scale=0.5]{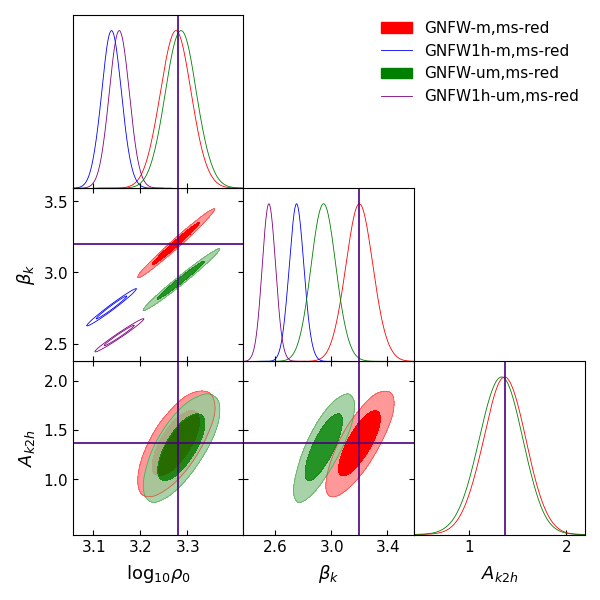}
	\vspace{-0.3cm}
	\caption{\label{fig:corner2d_fiducial} Comparison of parameter spaces for fits of the 2D projections ms-red sample with the GNFW-\textit{m} model as the fiducial model. The solid lines indicate the values of the fits of the 3D profiles for the same sample (ms-red, GNFW-\textit{m}). This figure shows that for density the most important modeling choice of the fitting model is the inclusion of a two-halo term, and for pressure both inclusion of distribution weighting and two-halo term are important.}
\end{figure*}

Figure~\ref{fig:corner2d_fiducial} shows contours only for the ms-red sample, but with different fitting models and distribution-matching options. The solid lines show the corresponding value of the 3D best fit for each parameter, which is used as input for the 2D fits. This figure differs from Figure~\ref{fig:corner2d_samples} in that the fits all use the same input profile $\vec{d}$ for the 2D observable profiles, calculated by the GNFW-\textit{m} model, but use different fitting models ($\vec{\mu}$), such as without a two-halo term or without mass-distribution matching. By treating the GNFW-\textit{m} profile as our truth, we can show how well the other models recover the given parameters. Simply by looking at the contours we can see that assuming an incorrect model results in estimates of the parameters not being very close to the actual values. For density, the GNFW-um model returns nearly correct values for $\log_{10}\rho_0$ and $A_{k2h}$, although slightly lower values for $\beta_k$, indicating that inclusion of distribution matching is not as important to kSZ model fitting as the inclusion of a two-halo term. The estimates from the fits with GNFW1h models are much lower than the true values for $\log_{10}\rho_0$ and $\beta_k$, with no information on a two-halo term.

For pressure we see more of a spread in returned values of different fitting models. The model that returns the closest value to the truth for $P_0$ is the GNFW1h-um model, for $\beta_t$ the GNFW1h-\textit{m} model, and for $A_{t2h}$ the GNFW-um model returns nearly the correct value. This indicates that the inclusion of mass-distribution matching is more important for tSZ model fitting than kSZ, and is equally as important as the inclusion of a two-halo term (see the $\sigma$ values of Table~\ref{tab:fits_2d}). For both of the profiles, incorrectly assuming the model induces systematic effects in the resulting parameters.

The resulting values of the 2D fits with the fiducial ms-red, GNFW-\textit{m} model as the data are shown in Table~\ref{tab:fits_2d} (contours shown in Figure~\ref{fig:corner2d_fiducial}), allowing for comparisons of the importance of different modeling choices. Next to the estimates for each parameter in the table is the number of $\sigma$ away from the corresponding fiducial value of the GNFW-\textit{m} model.

As a specific example, in the right panel of Figure~\ref{fig:corner2d_fiducial} we show contours of the ms-red sample for density. Taking the GNFW-\textit{m} model as our truth, the input values for $\log_{10}\rho_0$, $\beta_k$, and $A_{k2h}$ were $3.28\pm0.04$, $3.20\pm0.11$, and $1.37\pm0.23$, respectively. If we fit this profile with the GNFW-um model, we would get values for these parameters of $3.29\pm0.04$, $2.95\pm0.10$, and $1.33\pm0.23$, which would be differences of $0.2\sigma$, $1.7\sigma$, and $0.1\sigma$. If we fit the profile with the GNFW1h-\textit{m} model, we would get values for $\log_{10}\rho_0$ and $\beta_k$ of $3.12\pm0.02$ and $2.71\pm0.06$, which are differences of $3.6\sigma$ and $3.9\sigma$, with no information on the two-halo term. Lastly, if we fit the profile with the GNFW1h-um model, we would get values for $\log_{10}\rho_0$ and $\beta_k$ of $3.14\pm0.02$ and $2.52\pm0.05$, differences of $3.1\sigma$ and $5.6\sigma$, respectively, also with no information on the two-halo term. 
For this sample, it is clear that the inclusion of a two-halo term to the fitting model is the most important modeling systematic, as the differences between the input and selected values for the GNFW-um model are smaller than the differences of the GNFW1h models. 

\section{Conclusions}
In summary, we have developed methods to extract halo information from the IllustrisTNG simulations. We derive halo samples of different properties of interest, such as within a specific mass range or selected by color, and stack the thermodynamic properties of density and pressure to create 3D radial profiles using \texttt{Illstack}. We fit these profiles to a GNFW profile, testing the importance of the halo sample modeling uncertainties of mass selection, color selection, mass-distribution matching, and addition of a two-halo term to the fitting model. We project the profiles into a 2D observing space using \texttt{Mop-c-GT}, resulting in profiles with units of CMB temperature shifts due to the kSZ and tSZ effects, and perform the same fitting routine. We test the importance of the different modeling uncertainties in 2D by modeling how SO would detect an observed sample (subsample of CMASS, analyzed in \citet{Amodeo2021}) of differently-derived halo populations. Below we reiterate the important takeaways of the study.

We show in several ways that mass selection, color selection, mass-distribution matching, and fitting to a model that includes the addition of a two-halo term having varying levels of importance when modeling a sample of halos. These results are seen by the differences in the theoretical 3D profiles (see Figure~\ref{fig:fits3d}) and the observational 2D profiles (see Figure~\ref{fig:projections}). The main trends of these figures are: 1. the stellar mass-selected samples tend to have higher density and pressure values than the halo mass-selected samples due to the shape of TNG's SHMR, and 2. weighting by the observed mass distribution has a larger effect on the pressure profiles due to its nonlinear dependence on mass, but is still relevant for the density profiles. 

To further demonstrate the importance of matching the observed mass distribution, we show that some GNFW parameters fit to the 3D theoretical profiles show dependence on mass, shown in Figure~\ref{fig:fits_3d_mass}. Therefore, modeling these mass trends by weighting to the observed mass distribution is important when modeling observed samples for cross correlations. 

For the redshift and mass ranges of the observed CMASS sample, we show that the fits are not sensitive to the properties of the inner/intermediate profiles for both pressure and density. With the error bars that are forecasted for SO, the inclusion of such inner parameters, especially in pressure, results in strong degeneracies with other parameters, shown in Figure~\ref{fig:fixedparams}. After fixing an additional parameter in the GNFW fits, we show that we are able to better constrain the remaining free parameters and better understand the fits in relation to the modeling uncertainties of interest. We show again that different selections in mass and color return either clear separations between populations or a spread of estimates for each parameter.

Lastly, in Figure~\ref{fig:corner2d_fiducial} we show in more detail that fitting to a model including weighting by the mass distribution and addition of a two-halo term are extremely important. Given a fiducial profile, not including either of these modeling choices returns incorrect values for each parameter. Shown by Table~\ref{tab:fits_2d}, for this sample the inclusion of a two-halo term tends to be more important than the inclusion of weighting by the mass distribution for the density profiles, seen by the lower $\sigma$ values from the fiducial profile's parameters. For the pressure profiles the $\sigma$ values for the models lacking matching by the mass distribution and a two-halo term are comparable, indicating that both of these systematics need to be addressed.

The uncertainties explored in this study are by no means an exhaustive list to consider when modeling a sample of halos. For example, as briefly discussed previously, measurements of masses on both sides of the SHMR are quite uncertain but in particular the measurements of stellar masses introduce significantly more systematic uncertainties. One could perform the same kind of study for this uncertainty by designing samples corresponding to different methods of estimating an observed sample's stellar mass, e.g. converting luminosity to stellar mass, and quantifying how the 3D and 2D profiles differ accordingly. We did explore the SHMR used in CMASS with respect to TNG by selecting and weighting halos according to either their halo masses and stellar masses. In Figure~\ref{fig:corner2d_samples} we can see the differences in the inferred profile parameters for both the density and pressure. Thus, going forward detailed modeling of the SHMR will be critical. Along similar lines the halo occupation distribution (HOD) is another modeling concern for future SZ observations \citep{Pandey2020}.

In the future, we plan to expand these studies to explore the prescriptions of different parameters used in simulations other than TNG, and study any potential trends with redshift. To expand the variety of simulations, we plan to use the Cosmology and Astrophysics with MachinE Learning Simulations (CAMELS) suite of simulations \citep{Camels2021}. CAMELS is a suite of thousands of simulations varying astrophysical and cosmological parameters, and will be very useful in further studying modeling choices such as the ones discussed in this paper. 

With a greater understanding of how to correctly model an observed sample, including the choices discussed in this study and more, we will be more prepared to interpret observations with ever-increasing signal-to-noise and resolution. Better interpretations of observations will allow for a better understanding of the current unsolved theoretical questions of the CGM and galaxy evolution as a whole. 

\acknowledgments{E.M. and N.B. acknowledge support from NSF grant AST-1910021. N.B. acknowledges support from the Research and Technology Development fund at the Jet Propulsion Laboratory through the project entitled ``Mapping the Baryonic Majority.'' E.S. is supported by the Chamberlain fellowship at Lawrence Berkeley National Laboratory. S.F. is supported by the Physics Division of Lawrence Berkeley National Laboratory.}

\bibliographystyle{aasjournal}
\bibliography{cits.bib}

\end{document}